\documentstyle[sprocl]{article}

\input{psfig}

\def\Journal#1#2#3#4{{#1} {\bf #2}, #3 (#4)}

\def\NPB{{\em Nucl. Phys.} B}
\def\PLB{{\em Phys. Lett.} B}
\def\PRL{\em Phys. Rev. Lett.}
\def\PRD{{\em Phys. Rev.} D}

\def\be{\begin{equation}}
\def\ee{\end{equation}}
\def\bea{\begin{eqnarray}}
\def\eea{\end{eqnarray}}

\begin{document}

\vspace*{-1.7cm}

\begin{flushright}
\begin{tabular}{l}
NEAP-55, November 1998
\end{tabular}
\end{flushright}

\bigskip
\title{QUANTUM ANALYSIS OF RYDBERG ATOM CAVITY DETECTOR
FOR DARK MATTER AXION SEARCH}

\author{K. YAMAMOTO}

\address{Department of Nuclear Engineering, Kyoto University, \\
Kyoto 606-8501, Japan
\\ E-mail: yamamoto@nucleng.kyoto-u.ac.jp} 

\author{S. MATSUKI}

\address{Nuclear Science Division, \\
Institute for Chemical Research, Kyoto University, \\
Uji, Kyoto 611-0011, Japan
\\E-mail: matsuki@carrack.kuicr.kyoto-u.ac.jp}

\maketitle\abstracts{
Quantum analysis is made on the dynamical system consisting
of the cosmic axions, photons and Rydberg atoms which are interacting
in the resonant cavity.
The atomic motion in a continuous incident beam is taken into account
properly in order to make a precise estimate
of the efficiency of the Rydberg atom cavity detector
for dark matter axion search.
}

\section{Rydberg atom cavity detector of axions}
The ``invisible" axions
\cite{Peccei77,WW,KSZD}
in the mass range $ m_a = 10^{-6}{\rm eV} - 10^{-3}{\rm eV} $
is one of the most promising candidates of non-baryonic dark matter.
\cite{Kolb90,Smith90,Raffelt90}.
The basic idea for the dark matter axion search is to convert axions
into microwave photons in a resonant cavity under a strong magnetic field
via Primakoff process, as originally proposed by Sikivie
\cite{Sikivie83,Krauss85}.
After the pioneering searches with amplification-heterodyne-method
\cite{dePanfillis87,Wuensch89,Hagmann90},
an advanced experiment by the US group is currently continued,
and some results have been reported recently,
where the KSVZ axion of mass $ 2.9 \times 10^{-6} $
to $ 3.3 \times 10^{-6} {\rm eV} $ is excluded at the 90\% confidence level
as the dark matter in the halo of our galaxy
\cite{Hagmann98}.
\footnote{Invited talk presented at the IDM'98, Buxton, UK,
7-11 September, 1998}

We here describe a quite efficient scheme for the axion search,
where Rydberg atoms are utilized to detect the axion-converted photons
\cite{MY91,OMY96}.
The scheme of the detection of axions with Rydberg atoms
via Primakoff process is as follows:
\begin{equation}
{\bf [~axion~]}
\begin{array}{c} B_0 \\ \Longleftrightarrow \\ { } \end{array}
{\bf [~photon~]}~ \! \!
\begin{array}{l}
- \! \! \! - n^\prime \\ \hspace{.15cm} \Updownarrow \\ - \! \! \! - n
\end{array}
{\bf [~atom~]} ~.
\end{equation}
The axions are first converted into photons under a strong magnetic field
$ B_0 $ in the resonant cavity.
Then, by aborbing these axion-converted photons
Rydberg atoms are excited from the lower state $ | n \rangle $
to the upper state $ | n^\prime \rangle $,
where the atomic transition frequency $ \omega_b $ is tuned approximately
to the cavity resonant frequency $ \omega_c $.
The excited atoms are finally detected quite efficiently
with the selective field ionization method
\cite{Gallagher94}.
By cooling the resonant cavity down to about 10 mK,
the number of excited atoms due to the thermal photons can be
reduced sufficiently to obtain a significant signal-to-noise ratio.
Hence the Rydberg atom cavity detector is expected to be
quite efficient for the dark matter axion search.

We here present a quantum analysis on the dynamical system consisting
of the axions, photons and Rydberg atoms
which are interacting in the resonant cavity.
The motion and uniform distribution of the Rydberg atoms
in the incident beam is especially taken into account.
This analysis provides a precise estimate
of the efficiency of the present detection scheme
of the dark matter axions.
Detailed descriptions will be presented elsewhere.

\section{The axion-photon-atom system and interactions in the cavity}

The quantum system in the cavity consists of
the resonant photons, the coherent cosmic axions and the Rydberg atoms,
which is described in terms of interacting oscillators with dissipation.
The thermal photon number $ {\bar n}_c $
is determined by the cavity temperature $ T_c $.
Its damping rate $ \gamma_c = \omega_c / Q $ is given
by the quality factor $ Q $.
The coherent axions are normalized in a box of the de Broglie wave length
$ \lambda_a = 2 \pi \hbar / ( \beta_a m_a ) \sim 100 {\rm m} $
for $ m_a \sim 10^{-5} {\rm eV} $ and the mean velocity
$ \beta_a \sim 10^{-3} $ of the galactic axions.
The axion number is then given
by $ {\bar n}_a \simeq ( \rho_a / m_a ) \lambda_a^3 $
with the enegy density $ \rho_a $ of the axions.
The dissipation of the coherent axions is characterized
by their energy spread as $ \gamma_a \sim \beta_a^2 m_a $.
All the atoms are prepared at the lower state
with $ {\bar n}_b = 0 $.
Their dissipation may be neglected, $ \gamma_b \approx 0 $,
since the transitions to the off-resonant states are highly suppressed
in the resonant cavity.

The effective axion-photon coupling in the resonant cavity
with $ \omega_c \simeq m_a $ is calculated from the original lagrangian
\cite{OMY96} as
\begin{eqnarray}
\kappa & = & 4 \times 10^{-26} {\rm eV}
\left( g_{a \gamma \gamma} / 1.4 \times 10^{-15} {\rm GeV}^{-1} \right)
\left( G B_0 / 4 {\rm T} \right)
\nonumber \\
& \times &
\left( \beta_a m_a / 10^{-3} \times 10^{-5} {\rm eV} \right)^{3/2}
\left( V_1 / 5000 {\rm cm}^3 \right)^{1/2} ,
\end{eqnarray}
where $ V_1 $ and $ G $ are the volume and form factor
of the conversion cavity, respectively.
This coupling is estimated for the DFSZ axion.
The collective atom-photon coupling
\cite{Haroche85}
is gvien by
\begin{equation}
\Omega_N = 1 \times 10^{-10} {\rm eV}
\left( \Omega / 5 \times 10^3 {\rm s}^{-1} \right)
\left( N / 10^3 \right)^{1/2} ,
\end{equation}
where $ N $ is the number of atoms,
and the single atom-photon coupling
$ \Omega = ( d / \hbar )( \hbar \omega_c / 2 \epsilon_0 V_2 )^{1/2}
\alpha_2 $ is calculated with the electric dipole matrix element
$ d $ for the atomic transition and the electric field normalization
factor $ \alpha_2 $ ($ < 1 $) in the detection cavity with volume $ V_2 $.
This collective atom-photon coupling is comparable to
the photon damping rate
\begin{equation}
\gamma \equiv \gamma_c = 5 \times 10^{-10} {\rm eV}
\left( \omega_c / 10^{-5}{\rm eV} \right)
\left( 2 \times 10^4 / Q \right) .
\end{equation}

\section{Quantum evolution of the system}

The quatum evolution of the axion-photon-atom system is governed
by the Liouville equation with the total Hamiltonian
\begin{equation}
H = \sum_i \hbar \omega_i q_i^\dagger q_i
+ \hbar \kappa ( a^\dagger c + a c^\dagger )
+ \hbar \Omega_N ( b^\dagger c + b c^\dagger )
\end{equation}
for $ q_i $ = $ b $(atom), $ c $(photon) and $ a $(axion).
Then, the master equation for the second order moments
$ {\cal N}_{ij} = {\rm Tr}[ q_i^\dagger q_j \rho ] $
is derived by the usual procedure
\cite{Louisell90}:
\begin{equation}
\frac{d{\cal N}}{dt} = -i {\cal N} {\cal H}^T + i {\cal H}^* {\cal N}
+ {\cal D} ~,
\end{equation}
where
\begin{equation}
{\cal H} = \left( \begin{array}{ccc}
\Omega_b & \Omega_N & 0 \\
\Omega_N & \Omega_c & \kappa \\
0 & \kappa & \Omega_a  \\
\end{array} \right) ~,~~
{\cal D} = \left( \begin{array}{ccc}
\gamma_b {\bar n}_b & 0 & 0 \\
0 & \gamma_c {\bar n}_c & 0 \\
0 & 0 & \gamma_a {\bar n}_a
\end{array} \right)
\end{equation}
with $ \Omega_i \equiv \omega_i - i ( \gamma_i / 2 ) $.

The master equation is readily solved
with a initial condition $ {\cal N}_{ij} (0) = \delta_{ij} {\bar n}_i $.
(The atoms are prepared in the lower state with $ {\bar n}_b = 0 $.)
In particular, the number of excited atoms is given by
\begin{equation}
n_b (t) = {\cal N}_{bb} (t)
= r_{bc}(t) {\bar n}_c + r_{ba}(t) {\bar n}_a ~,
\end{equation}
which is expressed as the sum of the contributions
of the thermal backgroud photons and the axion-converted photons.
The axion-converted photon signal appears with the tiny factor
$ r_{ba} \propto ( \kappa / \gamma )^2 $,
which however can be compensated by the huge galactic axion
number $ {\bar n}_a $.

\section{Estimate of detection efficiency}

We make the quantum analysis in the following way
by taking into account
(i) the atomic motion and (ii) the almost uniform distribution
of the atoms in the incident beam:
The electric field felt by the atoms
varies with time through the atomic motion of velocity $ v $.
Accordingly, the atom-photon coupling becomes time-dependent.
In order to treat the atomic distribution,
we divide the atoms in the cavity into $ K $ bunches
for the fixed beam intensity $ I_{\rm Ryd} = N/t_{\rm tr} $,
where $ N $ is the total number of Rydberg atoms in the cavity,
and $ t_{\rm tr} $ is the atomic transit time through the cavity.
Then, the coutinuous beam is realized for $ K \gg 1 $.
The collective atomic mode of each bunch is denoted by $ b_i $
($ i = 1,2, \ldots , K $).
Then, the effective Hamiltonian $ {\cal H} $ becomes
a $ (K+2) \times (K+2) $ matirix.

Suppose that the $ i $-th atomic bunch locates around
$ x_i = (i-1) \Delta x $
for a time interval $ t_0 \leq t < t_0 + t_{\rm tr}/K $.
Here, $ \Delta x = L/K $ is the space interval of the atomic bunches
for the detection cavity length $ L $,
and the atomic transit time is given by $ t_{\rm tr} = L/v $,
which is typically $ \sim 100 \gamma^{-1} $.
Since the atomic number in each bunch is $ N/K $,
the collective atom-photon coupling is replaced as
\begin{equation}
\Omega_N \rightarrow
\Omega_i (t) = ( \Omega_N / {\sqrt K} ) f( x_i + v(t-t_0) ) ~,
\end{equation}
where $ f(x) $ is the electric field profile of the detection cavity.

After each time interval of $ t_{\rm tr}/K $,
the $ K $-th atomic bunch leaves the cavity,
and a new one enters.
Hence, the master equation of $ {\cal N} $ should be solved
with a suitable matching condition
at $ t_0 = M ( t_{\rm tr} / K ) $ ($ M = 1,2, \ldots $)
by substituting the collective modes
as $ b_i \rightarrow b_{i+1} $.
Then, by solving the master equation for many time intervals ($ M \gg 1 $)
the steady solution $ n_{b_i} (t) $ is realized,
which is the sum of the contributions of the axions
and thermal photons:
\begin{equation}
n_{b_i} (t) = n_{b_i}^a (t) + n_{b_i}^c (t) ~.
\end{equation}
The counting rates of the excited atoms
due to the axion-converted photons and thermal photons
are calculated, respectively, as
\begin{equation}
R_s = n_{b_K}^a ( t_{\rm tr} / K )/( t_{\rm tr} / K ) ~,~~
R_n = n_{b_K}^c ( t_{\rm tr} / K )/( t_{\rm tr} / K ) ~.
\end{equation}

\begin{figure}[htb]
\vspace{0cm}
\hspace{-.7cm}
\psfig{figure=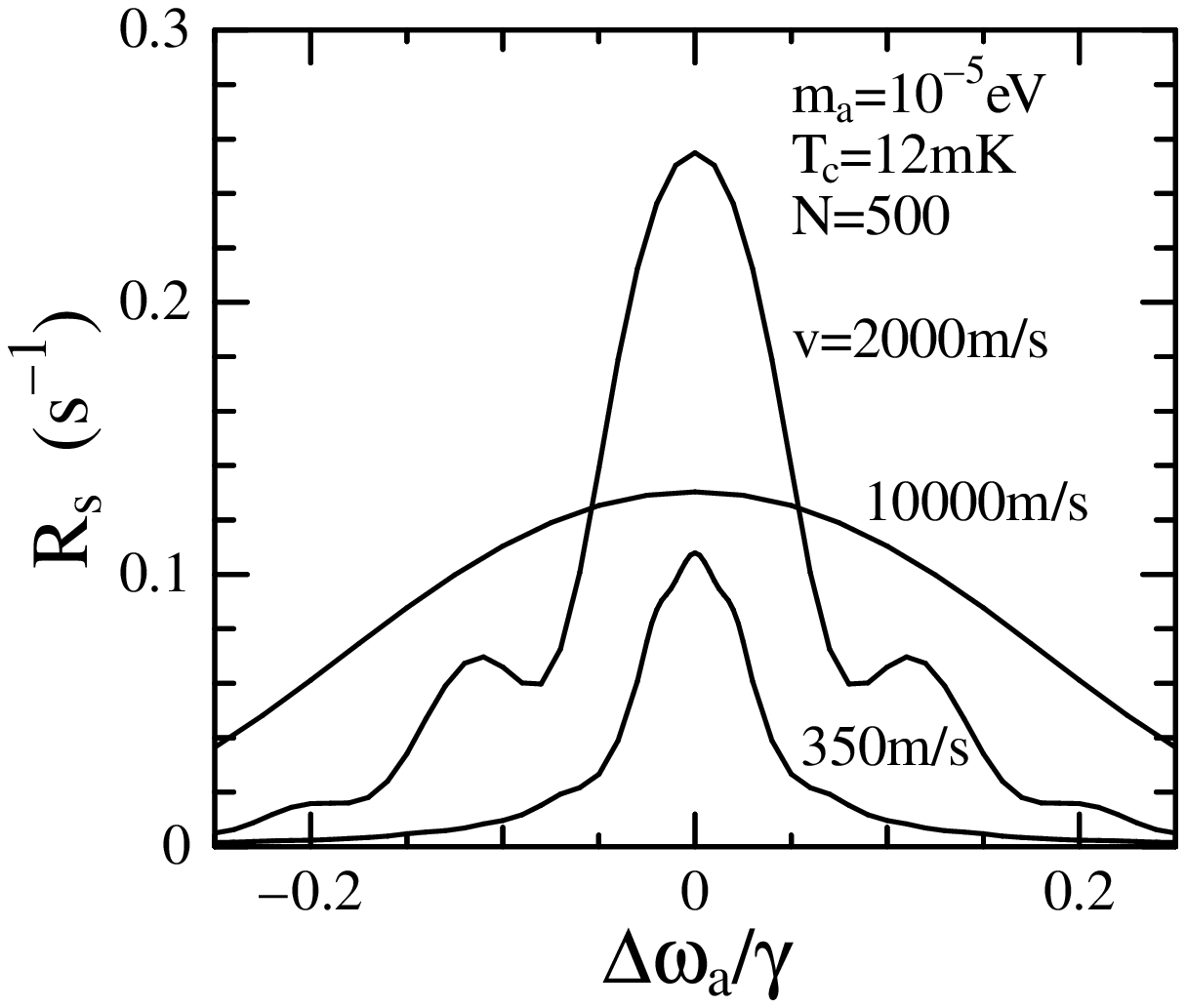,height=5cm}
\hspace{-.5cm}
\vspace{0cm}
\psfig{figure=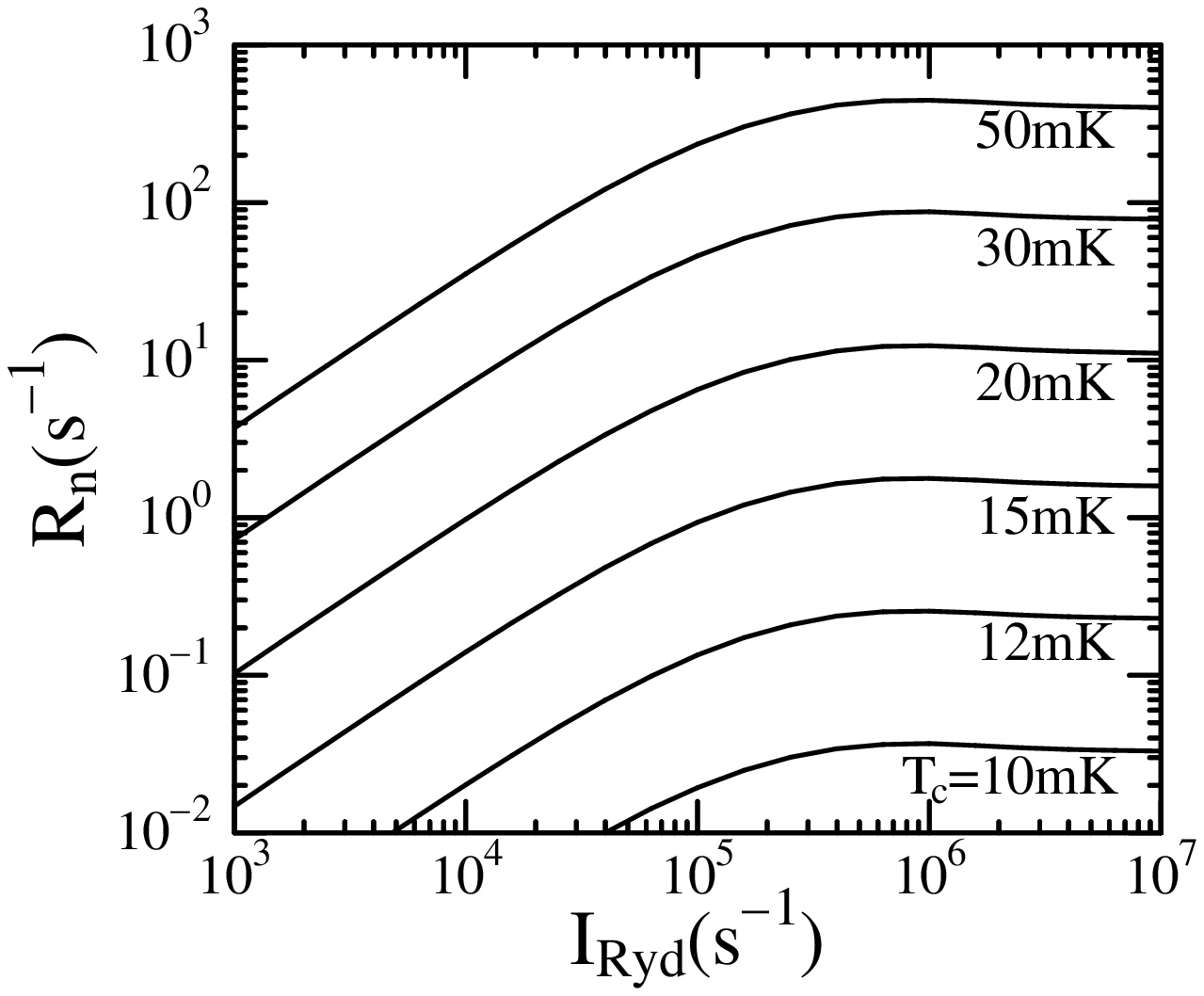,height=5cm}
\vspace{-.7cm}
\caption{The signal and noise rates.}
\vspace{-.1cm}
\label{fig:SN}
\end{figure}

Numerical calculations have been performed to estimate
the detection efficiency
for some practical values of the parameters such as
$ L = 0.2 {\rm m} $, $ Q = 2 \times 10^4 $,
$ T_c \sim 10{\rm mK} $,
$ \Omega = 5 \times 10^3 {\rm s}^{-1} $,
$ v = 350 {\rm m}/{\rm s} - 10000 {\rm m}/{\rm s} $,
$ I_{\rm Ryd} = 10^3 {\rm s}^{-1} - 10^7 {\rm s}^{-1} $
and $ \rho_a = \rho_{\rm halo} = 0.3 {\rm GeV}/{\rm cm}^3 $
giving $ {\bar n}_a = 6 \times 10^{25} $
for $ m_a = 10^{-5}{\rm eV} $ and $ \beta_a = 10^{-3} $.
The signal and noise rates are shown in Fig. \ref{fig:SN}.
\begin{figure}[htb]
\vspace{0cm}
\hspace{-.7cm}
\psfig{figure=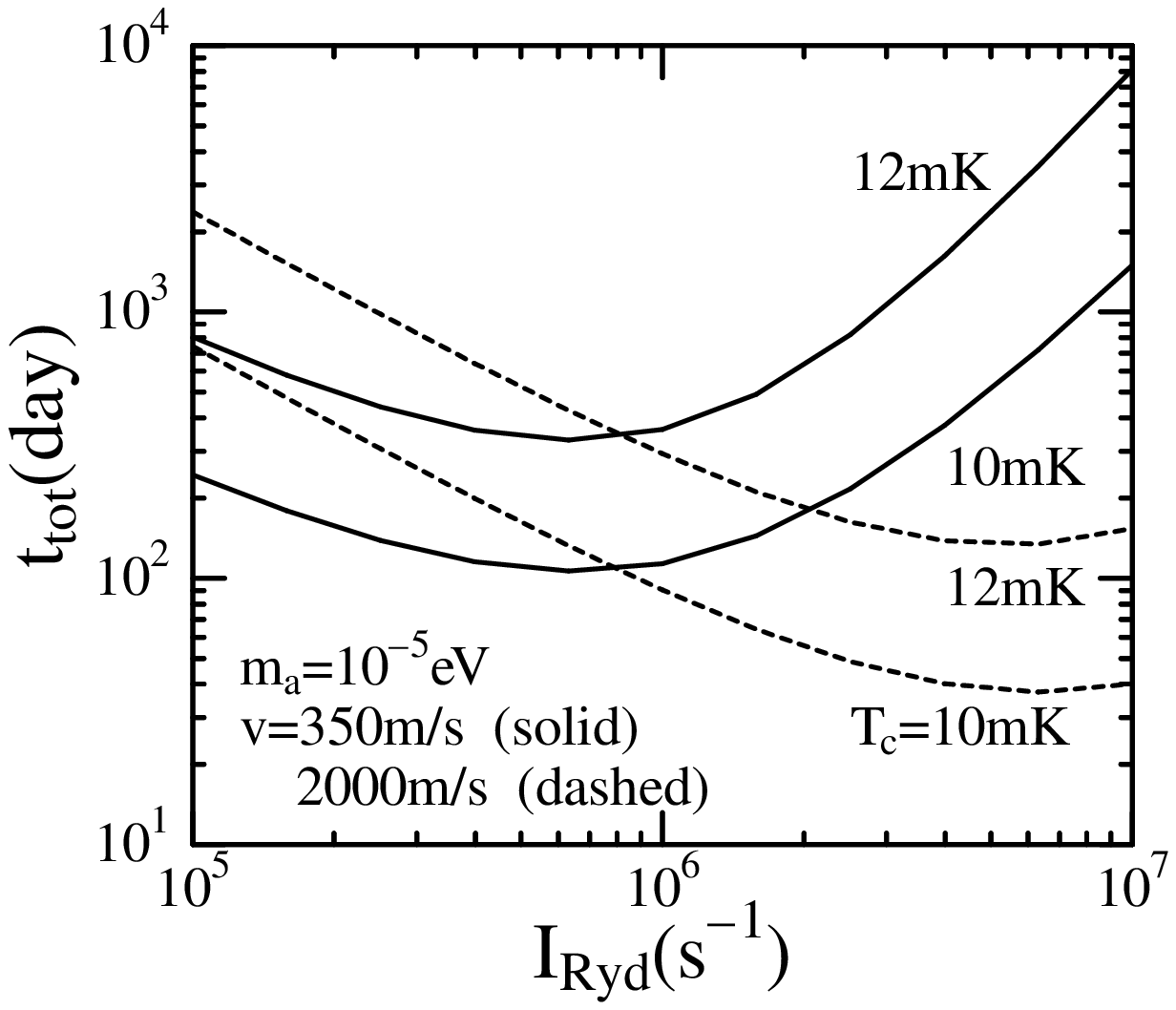,height=5cm}
\hspace{-.5cm}
\vspace{0cm}
\psfig{figure=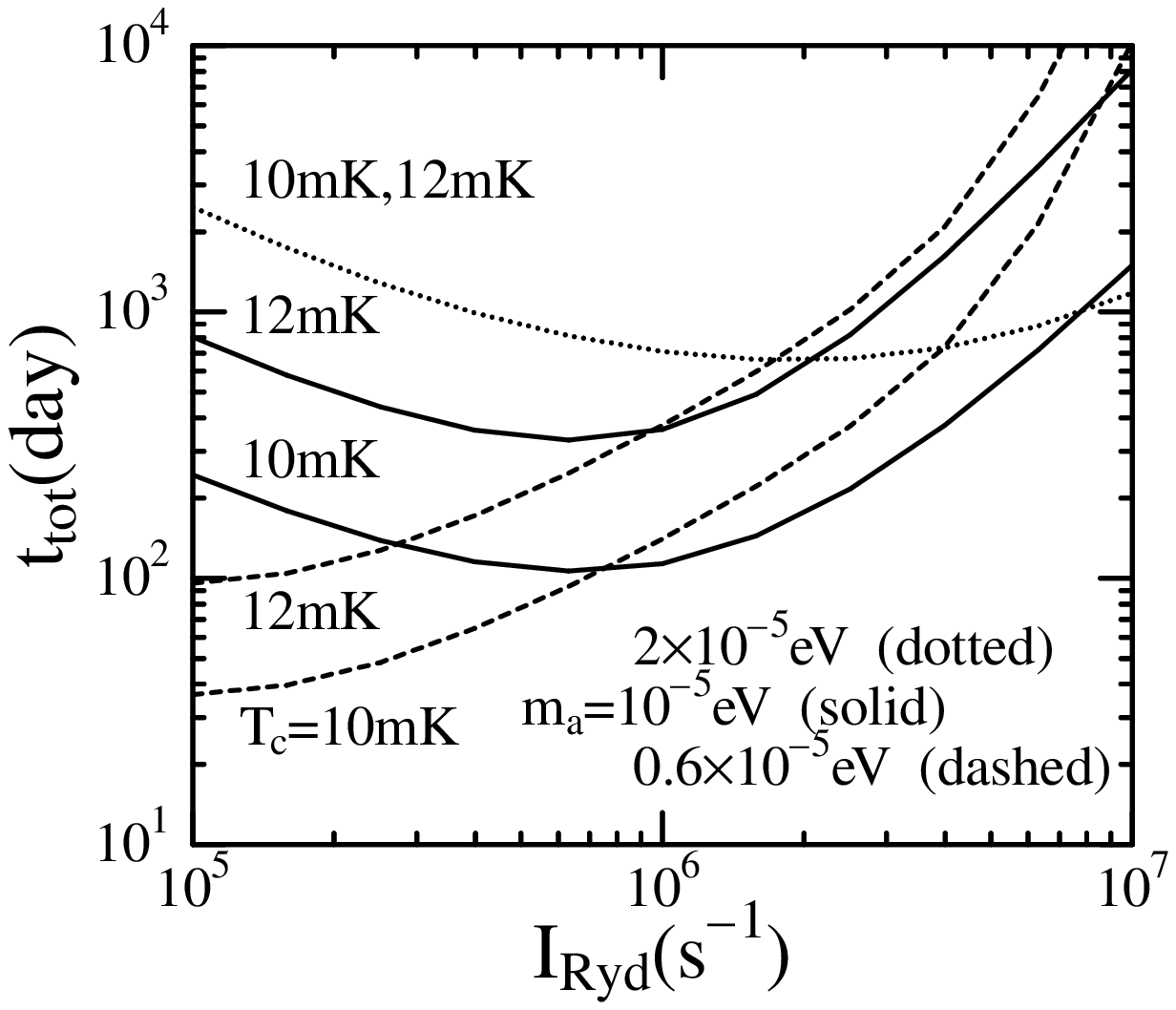,height=5cm}
\vspace{-.5cm}
\caption{The total scanning time depending on the atomic velocity $ v $
in the left and the axion mass $ m_a $
($ v = 350 {\rm m}/{\rm s} $) in the right.}
\vspace{-.3cm}
\label{fig:total}
\end{figure}
The signal form factor depending on the detuning of the axions
$ \Delta \omega_a \equiv \omega_a - \omega_c $
has a narrow width $ \sim 0.05 \gamma $
for $ v = 350 {\rm m}/{\rm s} $.
The noise rate, which is proportional to $ {\bar n}_c (T_c) $,
is estimated for various cavity temperatures
as a function of $ I_{\rm Ryd} = N/t_{\rm tr} $.
It is found that $ R_n $ is almost proportional
to $ \Omega_N^2 = \Omega^2 N \propto I_{\rm Ryd} $ for weak beam intensities.
On the other hand, for strong beam intensities the number of excited atoms
is saturated to certain values, where the atoms interact collectively
with the photons.

The measurement time required to search for the signal
at the confidence level of $ 3 \sigma $ is calculated by
\begin{equation}
\Delta t = 3^2 ( 1 + R_n / R_s )/R_s ~.
\end{equation}
The resonant condition for the absorption of the axion-converted photons,
as seen in Fig. \ref{fig:SN},
indicates that the frequency step of $ \omega_c $
($ \omega_b = \omega_c $) to be changed
in the axion search with unknown mass may be taken as
$ \Delta \omega_c = 0.05 \gamma $.
Then, the total scanning time over a 10\% frequency range
is given by
\begin{equation}
t_{\rm tot} = ( 0.1 \omega_c / \Delta \omega_c ) \Delta t ~.
\end{equation}
The estimate of $ t_{\rm tot} $ is shown in Fig. \ref{fig:total}.
We clearly find from this result that the DFSZ axion limit can be reached
for a reasonable scanning time with the Rydberg atom cavity detector.

\section{Summary}

We have developed quantum theoretical calculations
on the axion-photon-Rydberg system in the resonant cavity.
The analysis is made in the actual experimental situation
especially by taking into account
the motion and uniform distribution of Rydberg atoms
in the incident beam.
By using these calculations the detection efficiency is estimated,
which shows that the Rydberg atom cavity detector is quite efficient
for the dark matter axion search.

\section*{Acknowledgments}

This research was partly supported
by a Grant-in-Aid for Specially Promoted Research
by the Ministry of Education, Science, Sports and Culture,
Japan under the program No. 09102010.

\end{document}